\documentclass[12pt]{article}
\usepackage[textures]{graphics}
\parskip=\bigskipamount
\begin{document}
\title{\bf Characterisation of an entanglement-free evolution.}
\date{}
\author{}
\maketitle
\vglue -1.8truecm
\centerline{\large Thomas Durt\footnote{TONA Free University of
Brussels, Pleinlaan 2, B-1050 Brussels, Belgium. email: thomdurt@vub.ac.be}}\bigskip\bigskip

\noindent PACS number: O3.65.Bz

{\it Abstract: Two or more quantum systems are said to be in an entangled or non-factorisable
state if their joint (supposedly pure) wave-function is not expressible as a product of individual
wave functions but is instead a superposition of product states. It is only when the systems are
in a factorisable state that they can be considered to be separated (in the sense of Bell).
We show that whenever two quantum systems interact with each other, it is impossible that all
factorisable states remain factorisable during the interaction unless the full Hamiltonian does
not couple these systems so to say unless they do not really interact. We also present certain
conditions under which particular factorisable states remain factorisable although they represent a
bipartite system whose components mutually interact and pay a particular attention to the case where
the two particles interact mutually through an action at a distance in the three dimensional space.}

\section*{Introduction}

The term entanglement was first
introduced by Schroedinger who described this as the characteristic trait of quantum mechanics,
``the one that enforces its entire departure from classical
lines of thought'' [1]. Bell's inequalities [2] show that when two systems are prepared in an entangled
state, the knowledge of the whole cannot be reduced to the knowledge of the parts, and that to
some extent the systems lose their individuality. It is only when their joint wave-function is
factorisable that they are separable\footnote{It can be shown that whenever two distant systems are
in an entangled (pure) state, there exist well-chosen observables such that the associated
correlations do not admit a local realist explanation, which is revealed by the violation of
well-chosen Bell's inequalities [3,4].}. It is therefore interesting to investigate which are the
situations such that two systems, initially prepared in a (pure) product state remain in such a state although
they mutually interact.

We shall show (sections 1 and 2) that when the Hilbert spaces associated
to the interacting systems
$A$ and
$B$ are finite dimensional, if we impose that all the product states remain product states during
the interaction, the full Hamiltonian can be factorised as follows:
$H_{AB}(t)=H_{A}(t)\otimes I_{B}+I_{A}\otimes H_{B}(t)$, where
$H_{i}(t)$ acts on the ``$i$'' system only while $I_{j}$ is the identity operator on the ``$j$''
system ($i,j=A,B$). In other words, in quantum mechanics there is no interaction without
entanglement. We shall also present a sufficient condition under which particular factorisable
(non-necessary pure) states remain factorisable during the interaction.

We
shall discuss, in the section 3, the situation where the interacting systems are two
three-dimensional material points that interact through an action at a distance. We shall show that
the factorisability of the full wave-function is preserved (A) in the test-particle limit (when one
of the particles is quite more massive than the other one and is localised in a small region of space
during the interaction), (B) in the classical limit which is considered here to play relatively to
quantum wave mechanics a role comparable to the one played by geometrical optics relatively to
classical wave optics and (C) when the Hartree approximation is valid.
\section{Two interacting spin one-halve particles}
Firstly, let us consider the most simple situation: the systems $A$ and $B$ are spin one-halve
particles. We shall now show the following theorem: 

{\bf Theorem 0:}

Let us consider a system that consists of two spin one-halve particles $A$ and $B$. Let us assume that the
wave-function of the full system is a pure state of
${\bf C}^2\otimes {\bf C}^2$ which evolves according to Schroedinger's equation:

\begin{equation}\label{schrod}i
\hbar\,
\partial_t\, {\bf \Psi}_{AB}( t) = H_{AB}( t){\bf \Psi}_{AB}( t)\end{equation}Let us assume that
any arbitrary initially factorisable state ${\bf \Psi}_{AB}( t=0) =
\psi_{A}( t=0)\otimes\psi_{B}( t=0)$ remains factorisable during its temporal evolution:
${\bf \Psi}_{AB}( t) =
\psi_{A}( t)\otimes\psi_{B}( t)$ $\forall t \geq 0$.

Then, for each time $t\geq 0$, there exists a ``factorisable'' Hamiltonian
$H^{fact}_{AB}=H_{A}\otimes I_{B}+I_{A}\otimes H_{B}$ (where $H_{A(B)}$ is a
self-adjoint operator that acts on ${\bf C}^2$) which brings the same change at time $t$
as the change due to $H_{AB}( t)$.

{\bf Proof of the Theorem 0:}

Let us firstly consider that
initially the state of the system belongs to
a factorisable basis of ${\bf C}^2\otimes {\bf
C}^2$ that consists of the four following states: 
${\bf \Psi}^1_{AB}( t=0) = \big|+_A\big>\otimes\big|+'_B\big>$, ${\bf \Psi}^2_{AB}( t=0) =
\big|+_A\big>\otimes\big|-'_B\big>$,
${\bf \Psi}^3_{AB}( t=0) = \big|-_A\big>\otimes\big|+'_B\big>$, ${\bf \Psi}^4_{AB}( t=0) =
\big|-_A\big>\otimes\big|-'_B\big>$, where $\big|+(')_{A(B)}\big>$ and
$\big|-(')_{A(B)}\big>$ represent up and down spin states along conventional axes of
reference $Z_A(Z_B')$ assigned to the systems $A(B)$.

Necessarily, ${\bf \Psi}^1_{AB}( t) =
\big|\tilde +_A\big>\otimes\big|\tilde  +'_B\big>$ for some couple of directions $\tilde Z_A$,
$\tilde Z_B'$. The unitarity of the evolution law imposes that either ${\bf \Psi}^2_{AB}(
t) =
\big|\psi^2_{A}( t)\big>\otimes\big|\tilde -'_B\big>$ or ${\bf \Psi}^2_{AB}(
t) =
\big|\tilde -_A\big>\otimes\big|\psi^2_{B}( t)\big>$ where $\big|\psi^2_{A(B)}(
t)\big>$ is undetermined at this level of the proof. Let us consider now the first
alternative and assume that the system is initially prepared in the product state
${1
\over
\sqrt{2}}\cdot ({\bf \Psi}^1_{AB}( t) + {\bf \Psi}^2_{AB}( t))$ $=\ |+_A\big>\otimes {1 \over
\sqrt{2}}\cdot(\big|+'_B\big>+\big|-'_B\big>)$. In virtue of the linearity of the evolution law,
this state becomes at time $t$ the state  ${1
\over
\sqrt{2}}\cdot (\big|\tilde +_A\big>\otimes\big|\tilde 
+'_B\big> + \big|\psi^2_{A}( t)\big>\otimes\big|\tilde -'_B\big>)$ which is not a product state
unless
$\big|\psi^2_{A}( t)\big> = \big|\tilde +_A\big>$ up to a global phase-factor that we can consistently take to be
equal to unity (up to a redefinition of the phase of $\big|\tilde -'_B\big>$). Then,
${\bf
\Psi}^2_{AB}( t) =
\big|\tilde +_{A}\big>\otimes\big|\tilde -'_B\big>$. By a similar reasoning, the second alternative
leads to the conclusion that ${\bf \Psi}^2_{AB}(
t) =
\big|\tilde -_{A}\big>\otimes\big|\tilde +'_B\big>$. By repeating this proof with ${\bf \Psi}^3_{AB}$
instead of ${\bf \Psi}^2_{AB}$, we get that either
${\bf \Psi}^3_{AB}( t) =
\big|\tilde -_A\big>\otimes\big|\tilde +'_A\big>$ or ${\bf \Psi}^3_{AB}( t) =
\big|\tilde +_A\big>\otimes\big|\tilde -'_A\big>$. In virtue of the unitarity of the evolution
law, ${\bf \Psi}^3_{AB}( t)$ must be orthogonal to ${\bf \Psi}^2_{AB}( t)$ so that, in conclusion, two
alternatives remain possible: either (i) ${\bf \Psi}^2_{AB}( t) =
\big|\tilde +_{A}\big>\otimes\big|\tilde -'_B\big>$ and ${\bf \Psi}^3_{AB}(
t) =
\big|\tilde -_{A}\big>\otimes\big|\tilde +'_B\big>$ or (ii) ${\bf \Psi}^2_{AB}(
t) =
\big|\tilde -_{A}\big>\otimes\big|\tilde +'_B\big>$ and ${\bf \Psi}^3_{AB}(
t) =
\big|\tilde +_{A}\big>\otimes\big|\tilde -'_B\big>$. In any case, unitarity imposes that
${\bf \Psi}^4_{AB}( t) =
\big|\tilde -_{A}\big>\otimes\big|\tilde -'_B\big>$ up to a global phase. If moreover we require
that when the system is initially prepared in the product state
${1
\over
2}\cdot ({\bf \Psi}^1_{AB}( 0) + {\bf \Psi}^2_{AB}( 0) + {\bf \Psi}^3_{AB}( 0) + {\bf \Psi}^4_{AB}( 0))$ it remains in a
product state at time $t$, this global phase factor must be equal to unity. It is
easy to check that in both cases all states that are initially product states are still product
states at time
$t$. 

Let us consider firstly that the first alternative is valid. It is easy to find a ``factorisable''
Hamiltonian $H_{AB}^{fact}(t)=H_{A}(t)\otimes I_{B}+I_{A}\otimes H_{B}(t)$ that
sends ${\bf \Psi}^i_{AB}(
0)$ on ${\bf \Psi}^i_{AB}(
t)$ ($i = 1, 2, 3, 4)$ in a time $t$. Obviously, it is sufficient to choose $H_{A(B)}(t)$ in such a way that
$\big| +_{A(B)}\big>$ is sent onto $\big|\tilde
+(')_{A(B)}\big>$ in a time $t$. For instance we could take $H_{A(B)}(t)$ to be a multiple of a
time independent linear combination of the Pauli matrices that generates on the Bloch sphere a
rotation that brings
$\big| +_{A(B)}\big>$ onto $\big|\tilde
+(')_{A(B)}\big>$, and modulate the intensity of this Hamiltonian in order to perform the
rotation in a time
$t$. It is worth noting that such an Hamiltonian sends states that were initially product states on
product states for all intermediate times
$t'$ ($0 \leq t'\leq t$). Note that we could even let depend $H_{A}$ and $H_{B}$ on time and tailor them in
an ad-hoc way in order to generate arbitrary continuous state evolutions for all intermediate times
$t'$ ($0 \leq t'\leq t$) in so far the projections of $\big|
\tilde +_{A}(t')\big>$ and $\big|
\tilde +'_{B}(t')\big>$ are sufficiently regular curves that can be arbitrarily well approximated
by a series of arcs of circles on the Bloch sphere. However, in the present approach, time was discretised, and
it is not so simple to quantify properly what we mean by ``regular'', a limitation that we shall overcome in the
next section, in the proof of our main theorem, where the hypothesis of regularity in time is expressed quite
naturally by a requirement on the Taylor development of the temporal state evolution. 

Let us now consider the second alternative and assume that for all intermediate times $t'$ ($0
\leq t'\leq t$) the evolution sends states that were initially product states on product states.
Then, at time $t' = {t\over 2 }$ either the first alternative is valid or the second alternative
is valid. 

If the first alternative is valid at time $t'$ let us consider the time $t'' = {3t\over 4 }$. At
time $t''$ either the first alternative is valid or the second alternative is valid.
If the first alternative is valid let us consider the time $t''' = {7t\over 8 }$. Otherwise let us
consider the time $t''' = {5t\over 8 }$.

If the second alternative is valid at time $t'$ let us consider the time $t'' = {t\over 4 }$. At
time $t''$ either the first alternative is valid or the second alternative is valid.
If the first alternative is valid let us consider the time $t''' = {3t\over 8 }$. Otherwise let us
consider the time $t''' = {t\over 8 }$ and so on. 

By doing so it is easy to show that for any positive integer $N$ there must exist an intermediate time $t_0$ such
that the states
${\bf \Psi}^1_{AB}( t=t_0) = \big|+_A\big>_0\otimes\big|+'_B\big>_0$, ${\bf \Psi}^2_{AB}( t=t_0) =
\big|+_A\big>_0\otimes\big|-'_B\big>_0$,
${\bf \Psi}^3_{AB}( t=t_0) = \big|-_A\big>_0\otimes\big|+'_B\big>_0$, ${\bf \Psi}^4_{AB}( t=t_0) =
\big|-_A\big>_0\otimes\big|-'_B\big>_0$, (where $\big|+(')_{A(B)}\big>_0$ and
$\big|-(')_{A(B)}\big>_0$ represent up and down spin states along conventional axes of
reference $Z_A^0(Z_B^{0'})$ assigned to the systems $A(B)$) are sent at time $t_0 + \epsilon$ onto
the states ${\bf \Psi}^1_{AB}( t=t_0 + \epsilon) = \big|+_A\big>_\epsilon\otimes\big|+'_B\big>_\epsilon$,
${\bf \Psi}^2_{AB}( t=t_0 + \epsilon) =
\big|-_A\big>_\epsilon\otimes\big|+'_B\big>_\epsilon$,
${\bf \Psi}^3_{AB}( t=t_0 + \epsilon) = \big|+_A\big>_\epsilon\otimes\big|-'_B\big>_\epsilon$,
${\bf \Psi}^4_{AB}( t=t_0 +
\epsilon) =
\big|-_A\big>_\epsilon\otimes\big|-'_B\big>_\epsilon$, with $\epsilon = {t\over 2^N}$ (where
$\big|+(')_{A(B)}\big>_\epsilon$ and
$\big|-(')_{A(B)}\big>_\epsilon$ represent up and down spin states along conventional axes of
reference $Z_A^\epsilon(Z_B^{\epsilon'})$. For $N$ sufficiently large, the continuity of the
evolution law (eqn.\ref{schrod}) imposes that $\big|+_A\big>_\epsilon = \big|+_A\big>_0 + \tau
(\epsilon)$ and $\big|+_B\big>_\epsilon = \big|+_B\big>_0 + \tau
(\epsilon)$ where by definition $\tau
(\epsilon^m)$ decreases at least as fast as the $m$th power of $\epsilon$ when $\epsilon$ goes to
zero. But then the in-product between
${\bf \Psi}^2_{AB}( t=t_0)$ and
${\bf \Psi}^2_{AB}( t=t_\epsilon)$ is equal to zero (up to $\tau
(\epsilon)$). In virtue of Pythagoras's theorem their distance must be equal to $\sqrt 2$ (up to
$\tau (\epsilon)$) which contradicts the continuity of
the evolution law. Therefore, the second alternative is not valid and for any time $t$, we can
find a ``factorisable'' Hamiltonian $H_{AB}(t)=H_{A}(t)\otimes I_{B}+I_{A}\otimes H_{B}(t)$ that
sends ${\bf \Psi}^i_{AB}( 0)$ on ${\bf \Psi}^i_{AB}(
t)$ ($i = 1, 2, 3, 4)$ and sends product states on product states for all intermediate times.

Note that continuity in time plays a crucial role in our proof which is similar to the proof of the impossibility
of sending a righthand glove onto a lefthand glove by a continuous isometry of the three dimensional physical
(Euclidean) space, an intuitively obvious fact that presents deep analogies with the situation
encountered here. The transformation described in the second alternative can effectively be
obtained by composing the transformation described in the first alternative with a discrete
transformation during which the states of the systems $A$ and $B$ are interchanged. Similarly, the
composition of a continuous isometry (composition of Euclidean rotations and translations) and of a
reflection in a mirror sends a righthand glove on a ``virtual'', non-physical lefthand glove. It
is highly probable that we could refine the theorem 0 and generalise it to
arbitrary finite dimensional systems (where the full system is represented in the Hilbert
space ${\bf C}^{d_A}\otimes {\bf C}^{d_B}$, with $d_A$ and $d_B$ standing for the
dimensions of the systems $A$ and $B$) by following the same way of reasoning.
Nevertheless, we shall now give a general proof based on a different approach that is less
abstract but appeals more to physical intuition.
\section{Two interacting finite-dimensional systems}
We shall now present our main result.
Let us consider two interacting quantum systems $A$ and $B$. We assume that the Hilbert
spaces associated to these systems are finite dimensional (of dimensions $d_A$ and $d_B$
respectively), that the wave-function of the full system is a pure state of ${\bf
C}^{d_A}\otimes {\bf C}^{d_B}$ and obeys the Schroedinger equation: 
$i \hbar\, \partial_t\, {\bf \Psi}_{AB}( t) =
H_{AB}( t){\bf \Psi}_{AB}( t)$ where $H_{AB}( t)$ is a self-adjoint operator that acts on ${\bf
C}^{d_A}\otimes {\bf C}^{d_B}$, that we assume to be sufficiently regular in time in order to ensure
 that the temporal Taylor development of the
wave-function is valid up to the second order in time.

{\bf Main Theorem:}

All the product states
remain product states during the interaction if and only if the full Hamiltonian can be
factorised as follows:
\begin{equation}\label{fact}H_{AB}(t)=H_{A}(t)\otimes I_{B}+I_{A}\otimes H_{B}(t)\end{equation}where
$H_{i}$ acts on the $i$th system only while $I_{j}$ is the identity operator on the $j$th system
($i,j=A,B$).

In order to prove this theorem, we shall firstly prove the following lemma:

{\bf Lemma:}

A pure product state
remains product state during the interaction if and only if, during its evolution, the
Hamiltonian never couples this product state to a product state that is bi-orthogonal to it.

{\bf Proof of the Lemma:}

A) Proof of the necessary condition. Let us consider that at time $t$ the system is
prepared in a product state
${\bf
\Psi}_{AB}( t) =
\psi_{A}( t)\otimes\psi_{B}( t)$, and let us choose a
basis of product states $\big|\psi^i_{A}\big>\otimes \big|\phi^j_{B}\big>$ ($i: 1...d_A;
j:1...d_B$, and $\big<\psi^i_{A}\big|\psi^j_{A}\big> =
\delta_{ij}=\big<\phi^i_{B}\big|\phi^j_{B}\big>)$  such that $\psi_{A}( t) = \big|\psi^1_{A}\big>$
and
$\phi_{B}( t) = \big|\phi^1_{A}\big>$. Then, after a short time $\delta t$, \begin{displaymath}{\bf
\Psi}_{AB}( t+\delta t)= (I+{i\delta t\over \hbar }\cdot H_{AB}(t))\cdot {\bf \Psi}_{AB}( t)+
\tau(\delta t^2)\end{displaymath}where $\tau
(\epsilon^m)$ was defined in the previous section. In a matricial form, the previous equation
becomes:
\begin{equation}{\bf
\Psi}_{AB}( t+\delta t) =\big|\psi^1_{A}\big>\otimes\big|\phi^1_{B}\big>+{i\delta t\over \hbar
}\Sigma_{i:
1...d_A;j:1...d_B}H_{i1j1}\big|\psi^i_{A}\big>\otimes\big|\phi^j_{B}\big>+\tau(\delta
t^2)\end{equation} where
\begin{equation}\label{diantre} H_{ikjl}= 
\big<\psi^i_{A}\big|\otimes\big<\phi^j_{B}\big|H_{AB}(t)\big|\psi^k_{A}\big>\otimes\big|\phi^l_{B}\big>+
\tau(\delta t^2)\end{equation}
Equivalently,
\begin{displaymath}{\bf \Psi}_{AB}(
t+\delta t)= \big|\psi^1_{A}\big>\otimes\big|\phi^1_{B}\big>+{i\delta t\over \hbar}(\Sigma_{i:
1...d_A}H_{i111}\big|\psi^i_{A}\big>\otimes\big|\phi^1_{B}\big>\end{displaymath}
\begin{displaymath}+\Sigma_{j:
2...d_B}H_{11j1}\big|\psi^1_{A}\big>\otimes\big|\phi^j_{B}\big>+\Sigma_{i:
2...d_A;j:2...d_B}H_{i1j1}\big|\psi^i_{A}\big>\otimes\big|\phi^j_{B}\big> )+ \tau(\delta
t^2)\end{displaymath}All the components of ${\bf \Psi}_{AB}(
t+\delta t)$ that are bi-orthogonal to ${\bf
\Psi}_{AB}( t)$ are contained in the last term of
the previous equation: $\Sigma_{i:
2...d_A;j:2...d_B}H_{i1j1}\big|\psi^i_{A}\big>\otimes\big|\phi^j_{B}\big> )$, up to $\tau(\delta
t^2)$. We can
rewrite this equation as follows:
\begin{displaymath}{\bf \Psi}_{AB}(
t+\delta t)= (\big|\psi^1_{A}\big>+{i\delta t\over \hbar
}\Sigma_{i:
1...d_A}H_{i111}\big|\psi^i_{A}\big>)\otimes(\big|\phi^1_{B}\big>+{i\delta t\over \hbar
}\Sigma_{j:
2...d_B}H_{11j1}\big|\phi^j_{B}\big>)\end{displaymath}
\begin{equation}\label{equation}+{i\delta t\over \hbar
}\Sigma_{i:
2...d_A;j:2...d_B}H_{i1j1}\big|\psi^i_{A}\big>\otimes\big|\phi^j_{B}\big>+ \tau(\delta
t^2)\end{equation}
Let us assume that the Hamiltonian couples ${\bf \Psi}_{AB}(
t=0)$ to states that are bi-orthogonal to it, which means that $\Sigma_{i:
2...d_A;j:2...d_B}\big|H_{i1j1}\big|^2\not= 0.$ We shall now show that then the development of the
first order in $\delta t$ of the bi-orthogonal or Schmidt decomposition [5] of ${\bf \Psi}_{AB}(
t+\delta t)$ contains more than one product state, which means that ${\bf \Psi}_{AB}(
t+\delta t)$ is entangled for $\delta t$ small enough. In order to do so, let us consider the
components of
${\bf
\Psi}_{AB}( t+\delta t)$ that are bi-orthogonal to ${\bf
\Psi}_{AB}( t)$. In virtue of Schmidt's theorem of the bi-orthogonal decomposition [5], we can
find $d_A-1$ normalized states
$\big|\tilde
\psi^i_{A}\big>$ of
${\bf C}^{d_A}$ mutually orthogonal and orthogonal to $\big|
\psi^1_{A}\big>$ and $d_B-1$ normalized states $\big|\tilde \phi^j_{B}\big>$  of ${\bf
C}^{d_B}$ mutually orthogonal and orthogonal to $\big|
\phi^1_{B}\big>$ such that  \begin{displaymath}{i\delta t\over \hbar}\Sigma_{i:
2...d_A;j:2...d_B}H_{i1j1}\big|\psi^i_{A}\big>\otimes\big|\phi^j_{B}\big>=\Sigma_{i:
2...min(d_A,d_B)}\alpha_i \big|\tilde \psi^i_{A}\big>\otimes \big|\tilde
\phi^i_{B}\big>\end{displaymath}
Let us now define the state $\big|\tilde \psi^{1'}_{A}\big>$ of ${\bf C}^{d_A}$ as follows:
$\big|\tilde \psi^{1'}_{A}\big>={1\over N_1}\cdot (\big|\psi^1_{A}\big>+{i\delta t\over \hbar
}\Sigma_{i:
1...d_A}H_{i111}\big|\psi^i_{A}\big>)$, where $N_1$ is a normalisation factor, and let us replace
the orthonormal basis $\{\big| \psi^1_{A}\big>,\big|\tilde \psi^2_{A}\big>,\big|\tilde
\psi^3_{A}\big>,...,\big|\tilde \psi^{d_A}_{A}\big> \}$ of ${\bf C}^{d_A}$ by the orthonormal
basis $\{\big| \tilde \psi^{1'}_{A}\big>,\big|\tilde \psi^{2'}_{A}\big>,\big|\tilde
\psi^{3'}_{A}\big>,...,\big|\tilde \psi^{d_A'}_{A}\big> \}$ of ${\bf C}^{d_A}$ that we
obtain by the Gram-Schmidt orthonormalisation procedure:
\begin{displaymath}\big|\tilde \psi^{2'}_{A}\big>={1\over N_2}\cdot
(\big|\tilde \psi^2_{A}\big>- \big< \tilde \psi^{1'}_{A}\big|\tilde \psi^2_{A}\big>\cdot  \big|
\tilde
\psi^{1'}_{A}\big>)\end{displaymath}where $N_2$ is a normalisation factor.
\begin{displaymath}\big|\tilde \psi^{3'}_{A}\big>={1\over N_3}\cdot
(\big|\tilde \psi^3_{A}\big>- \big< \tilde \psi^{1'}_{A}\big|\tilde \psi^3_{A}\big>\cdot  \big|
\tilde
\psi^{1'}_{A}\big>- \big< \tilde \psi^{2'}_{A}\big|\tilde \psi^3_{A}\big>\cdot  \big|
\tilde
\psi^{2'}_{A}\big>)\end{displaymath}where $N_3$ is a normalisation factor, and so on. It is easy to
check that $\big|
\tilde
\psi^{i'}_{A}\big>=\big|
\tilde
\psi^i_{A}\big>+\tau(\delta t).$ Note that this is no longer true when the dimension $d_A$ is not
finite. We can repeat the same operation in order to replace
the orthonormal basis $\{\big| \psi^1_{B}\big>,\big|\tilde \psi^2_{B}\big>,\big|\tilde
\psi^3_{B}\big>,...,\big|\tilde \psi^{d_B}_{B}\big> \}$ of ${\bf C}^{d_B}$ by the orthonormal
basis $\{\big| \tilde \psi^{1'}_{B}\big>,\big|\tilde \psi^{2'}_{B}\big>,\big|\tilde
\psi^{3'}_{B}\big>,...,\big|\tilde \psi^{d_B'}_{b}\big> \}$ of ${\bf C}^{d_B}$. Then,
after substitition in the eqn.\ref{equation}, we obtain that: 
\begin{displaymath}{\bf \Psi}_{AB}(
t+\delta t)= \Sigma_{i:
1...min(d_A,d_B)}\alpha_i \big|\tilde \psi^{i'}_{A}\big>\otimes \big|\tilde
\phi^{i'}_{B}\big>+\tau(\delta t^2)\end{displaymath}where
\begin{equation}\label{schmidt}\big|\alpha_1\big|^2=1+\tau(\delta t^2), \Sigma_{i:
2...min(d_A,d_B)}\big|\alpha_i\big|^2 = {\delta t^2\over \hbar^2}\Sigma_{i:
2...d_A;j:2...d_B}\big|H_{i1j1}\big|^2+\tau(\delta t^3)\end{equation} The previous equation
expresses that the development up to the first order in $\delta t$ of the bi-orthogonal Schmidt
decomposition of ${\bf \Psi}_{AB}( t+\delta t)$ contains more than one product state. It is well
known that then ${\bf \Psi}_{AB}( t+\delta t)$ is an entangled state. Nevertheless, for those who
are not familiar with this property, we shall prove directly the result by making use of the
reduced density matrix. By definition, the reduced density matrix $\rho_A$ of the system $A$ is
equal to $Tr_B \rho$ where
$\rho$ is the projector on ${\bf \Psi}_{AB}$. Obviously, when the state of the system is a product
state (${\bf \Psi}_{AB} =\psi_{A}\otimes\psi_{B}$), $\rho_A$ is the projector on $\psi_{A}$, and
we have that $\rho_A$ = $\rho_A^2$, and $Tr\rho_A$ = $Tr\rho_A^2$ = 1. Actually, $Tr\rho_A$ -
$Tr\rho_A^2$ provides a good measure of the degree of the entanglement of the full system. If the
Schmidt bi-orthogonal decomposition of the state ${\bf \Psi}_{AB}$ is equal to $\Sigma_{i:
1...min(d_A,d_B)}\alpha'_i \big| \psi^{i'}_{A}\big>\otimes \big|
\phi^{i'}_{B}\big>$, then it is easy to check that $\rho_A = \Sigma_{i:
1...min(d_A,d_B)}\big|\alpha'_i \big|^2\big| \psi^{i'}_{A}\big>\big< \psi^{i'}_{A}\big|$, $Tr\rho_A =
\Sigma_{i: 1...min(d_A,d_B)}\big|\alpha'_i \big|^2=1$ by normalisation and $Tr\rho_A^2 =
\Sigma_{i: 1...min(d_A,d_B)}\big|\alpha'_i \big|^4\leq (Tr\rho_A)^2 = 1^2=1.$ The last inequality
is saturated for product states only. Note that $Tr\rho_A^2= Tr\rho_B^2$ which shows that this
parameter expresses properties of the system considered as a whole, as it must be when we are dealing
with entanglement. Obviously
$Tr\rho_A^2(t+\delta t)\leq
\big|\alpha_1
\big|^4+ (\Sigma_{i:
2...min(d_A,d_B)}\big|\alpha_i \big|^2)^2$. But $\big|\alpha_1
\big|^4 = (1-\Sigma_{i:
2...min(d_A,d_B)}\big|\alpha_i\big|^2)^2 = (1-{\delta t^2\over \hbar^2}\Sigma_{i:
2...d_A;j:2...d_B}\big|H_{i1j1}\big|^2+\tau(\delta t^3))^2$ and $(\Sigma_{i:
2...min(d_A,d_B)}\big|\alpha_i \big|^2)^2 = ({\delta t^2\over \hbar^2}\Sigma_{i:
2...d_A;j:2...d_B}\big|H_{i1j1}\big|^2+\tau(\delta t^3))^2$ in virtue of the eqn.\ref{schmidt}
so that $Tr\rho_A^2(t+\delta t) \leq 1-2 \cdot {\delta t^2\over \hbar^2}\Sigma_{i:
2...d_A;j:2...d_B}\big|H_{i1j1}\big|^2+\tau(\delta t^3)<1$ for $\delta t$ small
enough\footnote{It can be shown by direct computation that when the state of the system is a
product state (${\bf \Psi}_{AB}(t) =\psi_{A}(t)\otimes\psi_{B}(t)$), then the following
identity ${dTr\rho_A^2\over dt}(t)=0$ is necessarily satisfied, independently of the form of the
Hamiltonian $H_{AB}$. This explains why no term of the first order in time appears in the
previous development.}, which proves the necessary condition of the lemma.

B) Proof of the sufficient condition. Let us consider that at time $t$ the system is
prepared in a product state ${\bf \Psi}_{AB}( t) =
\psi_{A}( t)\otimes\psi_{B}( t)$, and let us choose a
basis of product states $\big|\psi^i_{A}\big>\otimes \big|\phi^j_{B}\big>$ similar to the basis
introduced in the proof of the necessary condition. When the Hamiltonian does not couple ${\bf
\Psi}_{AB}( t)$ to states that are bi-orthogonal to it,
$\Sigma_{i: 2...d_A;j:2...d_B}\big|H_{i1j1}\big|^2= 0$ (where $H_{ikjl}$ is defined in the
eqn.\ref{diantre}) and, in virtue of the eqn.\ref{equation}:
\begin{equation}i \hbar\, \partial_t\, {\bf \Psi}_{AB}( t) =
H_{AB}( t){\bf \Psi}_{AB}( t)= (\Sigma_{i:
1...d_A}H_{i111}\big|\psi^i_{A}\big>)\otimes
\big|\phi^1_{B}\big>+\big|\psi^1_{A}\big>\otimes(\Sigma_{j:
2...d_B}H_{11j1}\big|\phi^j_{B}\big>)\end{equation} We can rewrite this equation as follows:
\begin{equation}\label{eff3}i \hbar\, \partial_t\, {\bf \Psi}_{AB}( t) = (H^{eff.}_A( t)\cdot\psi_{A}(
t))\otimes\psi_{B}( t)+\psi_{A}( t)\otimes(H^{eff.}_B( t)\cdot \psi_{B}( t))\end{equation} 
where the effective Hamiltonians $H^{eff.}$ are defined as follows:
\begin{equation}\label{eff1}H^{eff.}_A(t)\cdot\rho_{A}( t)=Tr_B( H_{AB}(
t){\bf
\rho}_{AB}( t))\end{equation} and
\begin{equation}\label{eff2}H^{eff.}_B(t)\cdot
\rho_{B}( t)=Tr_A (H_{AB}( t){\bf \rho}_{AB}( t))-(Tr_{AB} (H_{AB}( t){\bf \rho}_{AB}(
t)))\cdot\rho_{B}( t)\end{equation} In these expressions $Tr_{i}$ represents the partial trace over
the degrees of freedom assigned to the system $i$ while 
${\bf
\rho}_{AB}( t)$ is the projector onto
${\bf
\Psi}_{AB}( t)$, $\rho_A(t)=Tr_B{\bf
\rho}_{AB}( t)$, and $\rho_B(t)=Tr_A{\bf
\rho}_{AB}( t)$. For instance, we have that \begin{displaymath}Tr_B( H_{AB}(
t){\bf
\rho}_{AB}( t))= \Sigma_{l: 1...d_B}\big<\phi^l_{B}\big| H_{AB}\big|\psi^1_{A}\big>\otimes
\big|\phi^1_{B}\big> \big<\psi^1_{A}\big|\otimes
\big<\phi^1_{B}\big| \phi^l_{B}\big>\end{displaymath}
\begin{displaymath} = \Sigma_{l:
1...d_B}\big<\phi^l_{B}\big| \Sigma_{i: 1...d_A,j:
1...d_B}H_{i1j1}\big|\psi^i_{A}\big>\otimes\big|\phi^j_{B}\big>
\delta_{l1}\big<\psi^1_{A}\big|\end{displaymath}

\begin{displaymath} = \Sigma_{l:
1...d_B} \Sigma_{i: 1...d_A,j:
1...d_B}H_{i1j1}\big|\psi^i_{A}\big>\delta_{lj}
\delta_{l1}\big<\psi^1_{A}\big|\end{displaymath}

\begin{displaymath}=(\Sigma_{i:
1...d_A}H_{i111}\big|\psi^i_{A}\big>\big<\psi^1_{A}\big|)\end{displaymath} so that $H^{eff.}_A( t)\cdot\psi_{A}(
t)=\Sigma_{i:
1...d_A}H_{i111}\big|\psi^i_{A}\big>$.

Let us consider the product state $\psi_{A}^{red}( t')\otimes\psi_{B}^{red}( t')$, where
$\psi^{red}_{A(B)}( t')$ is a solution of the reduced Schroedinger equation
$i
\hbar\,
\partial_{t'}\, 
\psi^{red}_{A(B)}( t') =
 H^{eff.}_{A(B)}( t')\cdot\psi^{red}_{A(B)}( t')$ for the initial condition $\psi^{red}_{A(B)}( t)$
=
$\psi_{A(B)}( t)$. Obviously, $i \hbar\, \partial_{t'}\,
\psi_{A}^{red}( t')\otimes\psi_{B}^{red}( t')=H_{AB}\psi_{A}^{red}( t')\otimes\psi_{B}^{red}( t')$
and
${\bf
\Psi}_{AB}( t) =
\psi_{A}^{red}( t)\otimes\psi_{B}^{red}( t)$ so that ${\bf \Psi}_{AB}( t') =
\psi_{A}^{red}( t')\otimes\psi_{B}^{red}( t')$, $\forall t' \geq t$ which ends the proof of the
lemma.

We shall now prove the main theorem.

{\bf Proof of the Main Theorem:}

A) Proof of the necessary condition. Let us choose a
basis of product states $\big|\psi^i_{A}\big>\otimes \big|\phi^j_{B}\big>$ ($i: 1...d_A;
j:1...d_B$ and $\big<\psi^i_{A}\big|\psi^j_{A}\big> =
\delta_{ij}=\big<\phi^i_{B}\big|\phi^j_{B}\big>)$.
If we impose that all the product states
remain product states during the interaction, then, in virtue of the lemma, the full
Hamiltonian never couples a product state to a product state that is bi-orthogonal to it. Then,
at any time $t$,
$\Sigma_{i: 2...d_A;j:2...d_B}\big|H_{i1j1}\big|^2= 0$ (where $H_{ikjl}$ is defined in the
eqn.\ref{diantre}) so that we have that:
\begin{displaymath}H_{AB}(t)\cdot\big|\psi^i_{A}\big>\otimes
\big|\phi^j_{B}\big>=\big|\triangle_A^{ij}\psi^i_{A}\big>\otimes\big|\phi^j_{B}\big>+
\big|\psi^i_{A}\big>
\otimes\big|\triangle_B^{ij}\phi^j_{B}\big>\end{displaymath}where
\begin{equation}\big|\triangle_A^{ij}\psi^i_{A}\big>=\Sigma_{k:
1...d_A}H_{kijj}\big|\psi^k_{A}\big>\end{equation} and
\begin{equation}\label{deltaB}\big|\triangle_B^{ij}\phi^i_{B}\big>=\Sigma_{k: 1...d_B, k\not=
j}H_{iikj}\big|\phi^k_{B}\big>\end{equation}

Let us consider that at time $t$ the system is prepared along one of the first four states ${\bf
\Psi}^i_{AB}$ ($i:1,...4$) of this basis: ${\bf \Psi}^1_{AB}( t) =
\big|\psi^1_{A}\big>\otimes\big|\phi^1_{B}\big>$, ${\bf \Psi}^2_{AB}( t) =
\big|\psi^1_{A}\big>\otimes\big|\phi^2_{B}\big>$,
${\bf \Psi}^3_{AB}( t) = \big|\psi^2_{A}\big>\otimes\big|\phi^1_{B}\big>$, ${\bf \Psi}^4_{AB}(
t) =
\big|\psi^2_{A}\big>\otimes\big|\phi^2_{A}\big>$. Then, 
\begin{displaymath}H_{AB}(t)\cdot{\bf
\Psi}^1_{AB}(t)=\big|\triangle_A^{11}\psi^1_{A}\big>\otimes\big|\phi^1_{B}\big>+\big|\psi^1_{A}
\big>\otimes \big|\triangle_B^{11}\phi^1_{B}\big>\end{displaymath}
\begin{displaymath}H_{AB}(t)\cdot{\bf
\Psi}^2_{AB}(t)=\big|\triangle_A^{12}\psi^1_{A}\big>\otimes\big|\phi^2_{B}\big>+\big|\psi^1_{A}
\big>\otimes \big|\triangle_B^{12}\phi^2_{B}\big>\end{displaymath}
\begin{displaymath}H_{AB}(t)\cdot{\bf
\Psi}^3_{AB}(t)=\big|\triangle_A^{21}\psi^2_{A}\big>\otimes\big|\phi^1_{B}\big>+\big|\psi^2_{A}
\big>\otimes \big|\triangle_B^{21}\phi^1_{B}\big>\end{displaymath}
\begin{displaymath}H_{AB}(t)\cdot{\bf
\Psi}^4_{AB}(t)=\big|\triangle_A^{22}\psi^2_{A}\big>\otimes\big|\phi2_{B}\big>+\big|\psi^2_{A}
\big>\otimes \big|\triangle_B^{22}\phi^2_{B}\big>\end{displaymath}
By linearity, 
\begin{displaymath}H_{AB}(t)\cdot {1\over \sqrt 2}({\bf
\Psi}^1_{AB}(t)+{\bf
\Psi}^3_{AB}(t))=H_{AB}(t)\cdot {1\over \sqrt
2}(\big|\psi^1_{A}\big>+\big|\psi^2_{A}\big>)\otimes\big|\phi^1_{B}\big>\end{displaymath}\begin{displaymath}={1\over
\sqrt
2}((\big|\triangle_A^{11}\psi^1_{A}\big>+\big|\triangle_A^{21}\psi^2_{A}\big>)\otimes\big|\phi^1_{B}\big>+\big|\psi^1_{A}
\big>\otimes \big|\triangle_B^{11}\phi^1_{B}\big>+\big|\psi^2_{A}
\big>\otimes \big|\triangle_B^{21}\phi^1_{B}\big>)\end{displaymath}
\begin{displaymath}={1\over
\sqrt
2}((\big|\triangle_A^{11}\psi^1_{A}\big>+\big|\triangle_A^{21}\psi^2_{A}\big>)\otimes\big|\phi^1_{B}\big>
+(\big|\psi^1_{A}\big>+\big|\psi^2_{A}\big>)\otimes
(\big|\triangle_B^{11}\phi^1_{B}\big>+\big|\triangle_B^{21}\phi^1_{B}\big>)\end{displaymath}
\begin{displaymath}+(\big|\psi^1_{A}\big>-\big|\psi^2_{A}\big>)\otimes
(\big|\triangle_B^{11}\phi^1_{B}\big>-\big|\triangle_B^{21}\phi^1_{B}\big>)\end{displaymath}
${1\over \sqrt 2}(\big|\psi^1_{A}\big>-\big|\psi^2_{A}\big>)$
is orthogonal to ${1\over \sqrt 2}(\big|\psi^1_{A}\big>+\big|\psi^2_{A}\big>)$, so that
$H_{AB}(t)\cdot {1\over \sqrt 2}({\bf
\Psi}^1_{AB}(t)+{\bf
\Psi}^3_{AB}(t))$ couples ${1\over \sqrt 2}({\bf
\Psi}^1_{AB}(t)+{\bf
\Psi}^3_{AB}(t))$ to a bi-orthogonal state unless
$(\big|\triangle_B^{11}\phi^1_{B}\big>-\big|\triangle_B^{21}\phi^1_{B}\big>)$ is parallel to
$\big|\phi^1_{B}\big>$. Now, ${1\over \sqrt 2}({\bf
\Psi}^1_{AB}(t)+{\bf
\Psi}^3_{AB}(t))$ is a product state so that, in virtue of the lemma, the following constraint
must be satisfied:
\begin{displaymath}(\big|\triangle_B^{11}\phi^1_{B}\big>-\big|\triangle_B^{21}
\phi^1_{B}\big>)=\lambda\big|\phi^1_{B}\big>\end{displaymath}
The same reasoning is valid with the states ${1\over \sqrt 2}({\bf
\Psi}^2_{AB}(t)+{\bf
\Psi}^4_{AB}(t))$, ${1\over \sqrt 2}({\bf
\Psi}^1_{AB}(t)+{\bf
\Psi}^2_{AB}(t))$ and ${1\over \sqrt 2}({\bf
\Psi}^3_{AB}(t)+{\bf
\Psi}^4_{AB}(t))$ and leads to the following constraints:
\begin{displaymath}(\big|\triangle_B^{12}\phi^2_{B}\big>-\big|\triangle_B^{22}
\phi^2_{B}\big>)=\lambda'\big|\phi^2_{B}\big>\end{displaymath}

\begin{displaymath}(\big|\triangle_A^{11}\psi^1_{A}\big>-\big|\triangle_A^{12}
\psi^1_{A}\big>)=\lambda''\big|\psi^1_{A}\big>\end{displaymath}

\begin{displaymath}(\big|\triangle_A^{21}\psi^2_{A}\big>-\big|\triangle_A^{22}
\psi^2_{A}\big>)=\lambda'''\big|\psi^2_{A}\big>\end{displaymath}

By definition (eqn.\ref{deltaB}), $\big|\triangle_B^{ij}\phi^j_{B}\big>$ is orthogonal to
$\big|\phi^j_{B}\big>$ so that necessarily $\lambda=\lambda'=0$.
Let us now consider the product state $({\bf
\Psi}^1_{AB}(t)+{\bf
\Psi}^2_{AB}(t)+{\bf
\Psi}^3_{AB}(t)+{\bf
\Psi}^4_{AB}(t))$.
By linearity:
\begin{displaymath}H_{AB}(t)\cdot {1\over  2}({\bf
\Psi}^1_{AB}(t)+{\bf
\Psi}^2_{AB}(t)+{\bf
\Psi}^3_{AB}(t)+{\bf
\Psi}^4_{AB}(t))=H_{AB}(t)\cdot {1\over 
2}(\big|\psi^1_{A}\big>+\big|\psi^2_{A}\big>)\otimes
(\big|\phi^1_{B}\big>+\big|\phi^2_{B}\big>)\end{displaymath}
\begin{displaymath}={1\over
\sqrt
2}((\big|\triangle_A^{11}\psi^1_{A}\big>+\big|\triangle_A^{21}\psi^2_{A}\big>)\otimes\big|\phi^1_{B}\big>
+(\big|\triangle_A^{12}\psi^1_{A}\big>+\big|\triangle_A^{22}\psi^2_{A}\big>)\otimes\big|\phi^2_{B}\big>\end{displaymath}
\begin{displaymath}
+\big|\psi^1_{A}
\big>\otimes (\big|\triangle_B^{11}\phi^1_{B}\big>+\big|\triangle_B^{12}\phi^2_{B}\big>)+\big|\psi^2_{A}
\big>\otimes (\big|\triangle_B^{21}\phi^1_{B}\big>+\big|\triangle_B^{22}\phi^2_{B}\big>))\end{displaymath}
In virtue of the constraints, we get that:
\begin{displaymath}H_{AB}(t)\cdot {1\over  2}({\bf
\Psi}^1_{AB}(t)+{\bf
\Psi}^2_{AB}(t)+{\bf
\Psi}^3_{AB}(t)+{\bf
\Psi}^4_{AB}(t))=\end{displaymath}
\begin{displaymath}={1\over
\sqrt
2}(\lambda''\big|\psi^1_{A}\big>+\lambda'''\big|\psi^2_{A}\big>)\otimes\big|\phi^1_{B}\big>
+(\big|\triangle_A^{12}\psi^1_{A}\big>+\big|\triangle_A^{22}\psi^2_{A}\big>)\otimes
(\big|\phi^1_{B}\big>+\big|\phi^2_{B}\big>)\end{displaymath}
\begin{displaymath}
+(\big|\psi^1_{A}\big>+\big|\psi^2_{A}\big>)
\otimes (\big|\triangle_B^{11}\phi^1_{B}\big>+\big|\triangle_B^{12}\phi^2_{B}\big>))\end{displaymath}
Such a state does not contain any state bi-orthogonal to ${1\over  2}({\bf
\Psi}^1_{AB}(t)+{\bf
\Psi}^2_{AB}(t)+{\bf
\Psi}^3_{AB}(t)+{\bf
\Psi}^4_{AB}(t))$ only if 
$\lambda''\big|\psi^1_{A}\big>+\lambda'''\big|\psi^2_{A}\big>=\lambda''''(\big|\psi^1_{A}\big>+\big|\psi^2_{A}\big>)$,
which imposes that $\lambda''\,=\,\lambda'''\,=\,\lambda''''$. We can repeat this proof with the indices $ii'$
for the system $A$ and $1j$ for the system $B$ instead of $12$ as it was the case in the previous proof, and
we obtain that
$\big|\triangle_B^{ij}\phi^j_{B}\big>=\big|\triangle_B^{i'j}\phi^j_{B}\big>=\big|\triangle_B^{j}\phi^j_{B}\big>$,
and
$\big|\triangle_A^{ij}\psi^i_{A}\big>=\big|\triangle_A^{i1}\psi^i_{A}\big>-
\lambda(j)\big|\psi^i_{A}\big>=\big|\triangle_A^{i}\psi^i_{A}\big>-\lambda(j)\big|\psi^i_{A}\big>$ (where
$\big|\triangle_A^{i}\psi^i_{A}\big>$ does not depend on $j$ while $\lambda(j)$ and
$\big|\triangle_B^{j}\phi^j_{B}\big>$ do not depend on
$i$). Therefore:
\begin{displaymath}H_{AB}(t)\cdot\big|\psi^i_{A}\big>\otimes
\big|\phi^j_{B}\big>=\big|\triangle_A^{i}\psi^i_{A}\big>\otimes\big|\phi^j_{B}\big>+
\big|\psi^i_{A}\big>
\otimes\big|\triangle_B^{j}\phi^j_{B}\big>-\lambda(j)\big|\psi^i_{A}\big>\otimes\big|\phi^j_{B}\big>\end{displaymath}
which fulfills the eqn.\ref{fact} provided we proceed to the following identifications:
$H_A(t)\cdot\big|\psi^i_{A}\big> =
\big|\triangle_A^{i}\psi^i_{A}\big>$ and $H_B(t)\cdot \big|\phi^j_{B}\big> =
\big|\triangle_B^{j}\phi^j_{B}\big>-\lambda(j)\big|\phi^j_{B}\big>$.
This ends the proof of the necessary condition of the main theorem.

B) Proof of the sufficient condition. Let us assume that the full Hamiltonian can be
factorised according to the eqn.\ref{fact}. Let us consider the product state $\psi_{A}^{red}(
t')\otimes\psi_{B}^{red}( t')$, where
$\psi^{red}_{A(B)}( t')$ is a solution of the reduced Schroedinger equation
$i
\hbar\,
\partial_{t'}\, 
\psi^{red}_{A(B)}( t') =
 H_{A(B)}( t')\cdot\psi^{red}_{A(B)}( t')$ for the initial condition $\psi^{red}_{A(B)}( t)$
=
$\psi_{A(B)}( t)$. Obviously, $i \hbar\, \partial_{t'}\,
\psi_{A}^{red}( t')\otimes\psi_{B}^{red}( t')=H_{AB}(t')\psi_{A}^{red}( t')\otimes\psi_{B}^{red}(
t')$ and
${\bf
\Psi}_{AB}( t) =
\psi_{A}^{red}( t)\otimes\psi_{B}^{red}( t)$ so that ${\bf \Psi}_{AB}( t') =
\psi_{A}^{red}( t')\otimes\psi_{B}^{red}( t')$, $\forall t' \geq t$ which ends the proof of the main
theorem.

Note that the condition \ref{fichtre} encountered in the proof of  the sufficient condition of the
lemma can be generalised to factorisable non-necessarily pure states. This is the essence of the
following theorem.

{\bf Theorem 2:}

If initially, a bipartite system is prepared in a factorisable (non-necessarily pure) state: ${\bf
\rho}_{AB}( t=0)=\rho_{A}( t=0)\otimes\rho_{B}( t=0)$, and that $\forall t \geq 0$\begin{equation}
\label{fichtre}H_{AB}( t){\bf
\rho}_{AB}( t)= (H^{eff.}_A( t)\cdot\rho_{A}(
t))\otimes\rho_{B}( t)+\rho_{A}( t)\otimes(H^{eff.}_B( t)\cdot \rho_{B}( t))\end{equation} 
where \begin{displaymath}H^{eff.}_A(t)\cdot\rho_{A}( t)=Tr_B( H_{AB}(
t){\bf
\rho}_{AB}( t))\end{displaymath} and
\begin{displaymath}H^{eff.}_B(t)\cdot
\rho_{B}( t)=Tr_A (H_{AB}( t){\bf \rho}_{AB}( t))-(Tr_{AB} (H_{AB}( t){\bf \rho}_{AB}(
t)))\cdot\rho_{B}( t),\end{displaymath}then, necessarily, the state remains factorisable during the interaction:
${\bf
\rho}_{AB}( t)=\rho_{A}( t)\otimes\rho_{B}( t)$ $\forall t \geq 0$.

{\bf Proof of the Theorem 2:}

When we describe the state of the system by a density matrix, its evolution obeys the von Neumann
equation:

\begin{equation}i \hbar\, \partial_t\, {\bf \rho}_{AB}( t) =
[ H_{AB}( t),{\bf \rho}_{AB}( t)]\end{equation}where $[X,Y]$ represents the commutator of two operators $X$ and
$Y$.
$H_{AB}(t)$ and
${\bf
\rho}_{AB}(t)$ are self-adjoint operators so that $[ H_{AB}( t),{\bf
\rho}_{AB}( t)]=  H_{AB}( t)\cdot{\bf
\rho}_{AB}( t)-( H_{AB}( t)\cdot{\bf \rho}_{AB}( t))^{+}$ where $O^{+}$ represents the self-adjoint operator of
$O$. Moreover, in virtue of the eqn.\ref{fichtre}, we have that:
\begin{displaymath}i \hbar\, \partial_t\, {\bf \rho}_{AB}( t) =
(H^{eff.}_A( t)\cdot\rho_{A}(
t))\otimes\rho_{B}( t)+\rho_{A}( t)\otimes(H^{eff.}_B( t)\cdot \rho_{B}( t))\end{displaymath} 
\begin{equation}-
((H^{eff.}_A( t)\cdot\rho_{A}(
t))\otimes\rho_{B}( t)+\rho_{A}( t)\otimes(H^{eff.}_B( t)\cdot \rho_{B}( t)))^+\end{equation} 

 Let us consider the product state $\rho_{A}^{red}( t)\otimes\rho_{B}^{red}( t)$, where $\rho^{red}_{A(B)}( t)$
is a solution of the reduced von Neumann equation $i \hbar\, \partial_t\, {\bf \rho}^{red}_{A(B)}( t) =
[ H^{eff.}_{A(B)}( t),{\bf \rho}^{red}_{A(B)}( t)]$ for the initial condition $\rho^{red}_{A(B)}( t=0)$ =
$\rho_{A(B)}( t=0)$. Obviously, $i \hbar\, \partial_t\,
\rho_{A}^{red}( t)\otimes\rho_{B}^{red}( t)=H_{AB}(t)\rho_{A}^{red}( t)\otimes\rho_{B}^{red}( t)$
and ${\bf
\rho}_{AB}( t=0) =
\rho_{A}^{red}( t=0)\otimes\rho_{B}^{red}( t=0)$ so that ${\bf \rho}_{AB}( t) =
\rho_{A}^{red}( t)\otimes\rho_{B}^{red}( t)$, $\forall t \geq 0$ which proves the
theorem\footnote{Note that this proof as well as the proof of the sufficient condition of the main
theorem are also valid when the systems
$A$ and
$B$ are infinite dimensional, for instance when they are localised particles that interact through a
central potential.}.

Note that the eqn.\ref{fichtre} is linear in the coupling Hamiltonian $H_{AB}$ and is automatically satisfied
when the eqn.\ref{fact} is satisfied. Nevertheless it is non-linear in ${\bf \rho}_{AB}$. Moreover,
the effective potential that acts onto say the $A$ particle is likely to depend on the state of the
$B$ particle, a situation that does not occur if we impose that all product states remain product
states. 

Beside, it is worth noting that the sufficient
condition expressed by the eqn.\ref{fichtre} is also necessary in the case of pure states.
Effectively, if, initially, the bipartite system is prepared in a factorisable pure state: ${\bf
\Psi}_{AB}( t=0) =
\psi_{A}( t=0)\otimes\psi_{B}( t=0)$, and that the state remains factorisable during the
interaction, then, in virtue of the necessary condition of the lemma, the Hamiltonian may not couple
the state
${\bf
\Psi}_{AB}( t)=\psi_{A}( t)\otimes\psi_{B}( t)$ at any time
$t \geq 0$ to a bi-orthogonal state so that, following the proof of the
lemma, the eqn.\ref{eff3} must be valid at any time. Therefore, in virtue of the eqn.\ref{schrod},
the eqn.\ref{fichtre} must be valid too.

Now, the sufficient condition
expressed by the eqn.\ref{fichtre} is in general not necessary in the case of non-pure states as
shows the following counterexample. If initially, the bipartite system is prepared in a factorisable
state:
${\bf
\rho}_{AB}( t=0)=\rho_{A}( t=0)\otimes\rho_{B}( t=0)$, and that $\forall t \geq 0, H_{AB}( t)={\bf
\rho}_{AB}( t=0)$, then it is easy to check that ${\bf
\rho}_{AB}( t=0)={\bf
\rho}_{AB}( t) \forall t \geq 0$, $ H^{eff.}_A(t)\cdot\rho_{A}( t)=Tr_B( H_{AB}(
t){\bf
\rho}_{AB}( t))= \rho_{A}^2( t=0)\cdot Tr_{B}\rho^2_{B}( t=0)$, $H^{eff.}_B(t)\cdot
\rho_{B}( t)=Tr_A (H_{AB}( t){\bf \rho}_{AB}( t))-(Tr_{AB} (H_{AB}( t){\bf \rho}_{AB}(
t)))\cdot\rho_{B}( t)=Tr_A\rho_{A}^2( t=0)\cdot \rho^2_{B}( t=0)-Tr_A\rho_{A}^2( t=0)\cdot Tr_{B}\rho^2_{B}(
t=0)\cdot \rho_{B}( t=0)$ and it is easy to check that in general the eqns. \ref{eff3} or
\ref{fichtre} are not valid when the initial state is not pure so to say when it is not a product of
pure states.

\section{The infinite dimensional case}
The proofs of the necessary conditions of the lemma (and thus of the
necessary condition of the main theorem) are not valid when the systems
$A$ and
$B$ are infinite dimensional. Nevertheless, we conjecture that these conditions are still true in
that case, so to say that there is no interaction without entanglement. Let us for instance consider
that
$A$ and $B$ are two distinguishable particles, and that their interaction potential is an action a
distance that is time-independent and invariant under spatial translations (a Coulombian interaction
for instance). They fulfill thus (in the non-relativistic regime) the following Schroedinger
equation:
\begin{displaymath}i \hbar\, \partial_t\, \Psi({\bf r}_{A},\, {\bf r}_{B},\, t) =
-({\hbar^2 \over 2m_A} \Delta_{A}\, +\,  {\hbar^2 \over 2m_B}\Delta_{B})\Psi({\bf r}_{A},\, {\bf
r}_{B},\, t)\end{displaymath}\begin{equation} +\, V_{AB} ({\bf r}_{A}-{\bf r}_{B})\Psi({\bf r}_{A},\, {\bf
r}_{B},\, t) \end{equation}
where $\Delta_{A(B)}$ is the Laplacian operator in the
$A(B)$ coordinates. As the potential does depend on the relative position ${\bf r}_{rel}={\bf r}_{A}-{\bf
r}_{B}$ only, it is convenient to pass to the center of mass coordinates:
\begin{displaymath}i \hbar\, \partial_t\, \Psi({\bf r}_{CM},\, {\bf r}_{rel},\, t) =
- ({\hbar^2 \over 2(m_A+m_B)}\Delta_{CM}\, +\,  {\hbar^2 \over 2\mu}\Delta_{rel})\Psi({\bf r}_{CM},\, {\bf
r}_{rel},\, t)\end{displaymath}\begin{equation} +\, V_{AB} ({\bf r}_{rel})\Psi({\bf r}_{CM},\, {\bf
r}_{rel},\, t) \end{equation}where ${\bf r}_{CM}={m_A{\bf r}_{A}+m_B{\bf r}_{B}\over m_A+m_B}$ and
$\mu={m_A\cdot m_B\over m_A+m_B}$. As it is well-known, the previous equation is separable which means that
if, initially, the wave-function is factorisable in these coordinates, it will remain so during the evolution.
Now, we are interested in situations for which the wave-function is initially factorisable according to the
partition of the Hilbert space that is induced by the systems $A$ and $B$. In general, such a wave-function is
not factorisable in the coordinates of the center of mass. Formally, if $\Psi({\bf r}_{A},\, {\bf r}_{B},\, t=0)
= \psi_A({\bf r}_{A},\, t=0)\cdot\psi_B({\bf r}_{B},\, t=0)$, $\Psi({\bf r}_{CM},\, {\bf r}_{rel},\, t=0)=\int
d\omega A(\omega) \psi^{\omega}_{CM}({\bf r}_{CM},\, t=0)\cdot\psi^{\omega}_{rel}({\bf r}_{rel},\, t=0)$ where
$A(\omega)$ is a generally non-peaked amplitude distribution. Then, at time $t$, $\Psi({\bf r}_{CM},\, {\bf
r}_{rel},\, t)=\int d\omega A(\omega)\psi^{\omega}_{CM}({\bf r}_{CM},\, t)\cdot\psi^{\omega}_{rel}({\bf
r}_{rel},\, t),$ where
$\psi^{\omega}_{CM}({\bf r}_{CM},\, t)$ obeyed during the time interval $[0,t]$ a free Schroedinger evolution
for the initial condition
$\psi^{\omega}_{CM}({\bf r}_{CM},\, t=0)$ while $\psi^{\omega}_{rel}({\bf r}_{rel},\, t)$ was submitted to the
interaction potential $V_{AB}(r_{rel})$. In general, $\Psi({\bf r}_{A},\, {\bf r}_{B},\, t)$ is no longer
factorisable into a product of the form $\psi_A({\bf r}_{A},\, t)\cdot\psi_B({\bf r}_{B},\, t)$. Actually,
this is not astonishing because, in virtue of Noether's theorem the full momentum is conserved during the
evolution. Therefore the recoil of one of the two particles could be used in order to determine (up to the
initial undeterminacy of the centre of mass) what is the recoil of the second particle. The existence of
such correlations is expressed by the entanglement of the full wave-function. On the basis of such general
considerations we expect that entanglement is very likely to occur due to the interaction between the two
particles. 

Nevertheless, it is interesting to investigate in which situations it is a good approximation to consider that
the systems $A$ and
$B$ remain in a factorisable state during time. We shall distinguish three typical situations.
\subsection{Scattering of a light particle by a heavy and well localized target (the test-particle limit)}
Let us assume that $m_A<<m_B$, and that the $B$ particle is initially at rest and well localized. The particle
$A$ is assumed to be initially prepared in such a way that it will pass in the vicinity of the heavy
particle $B$, that its trajectory will undergo a deviation due to the influence of the interaction $V_{AB}$,
and that it will finally escape to infinity without exerting any significant back action onto the particle $B$.
This situation is often encountered during scattering experiment. If we let coincide the origin of the system of
coordinates associated to the particle
$B$ with its location, and that we neglect its recoil as well as its dispersion (this approximation is only
valid during a limited period of time), the following approximations are valid: $ {\bf r}_{CM}\approx {\bf
r}_{B}\approx 0$, ${\bf r}_{rel}\approx {\bf r}_{A}-0 ={\bf r}_{A}$, $\psi_A({\bf r}_{A},\,
t)\approx\psi_{rel}({\bf r}_{rel},\, t)$ and $\psi_B({\bf r}_{B},\, t)\approx\psi_{CM}({\bf r}_{CM},\, t)$.
Moreover,
$\Psi({\bf r}_{A},\, {\bf r}_{B},\, t=0) = \psi_A({\bf r}_{A},\, t=0)\cdot\psi_B({\bf r}_{B},\, t=0)\approx
\psi_{rel}({\bf r}_{rel},\, t=0)\cdot\psi_{CM}({\bf r}_{CM},\, t=0)\approx\Psi({\bf r}_{CM},\, {\bf r}_{rel},\,
t=0)$. At time $t$, $\Psi({\bf r}_{CM},\, {\bf r}_{rel},\, t)\approx \psi_{rel}({\bf r}_{rel},\,
t=0)\cdot\psi_{CM}({\bf r}_{CM},\, t)\approx \psi_A({\bf r}_{A},\, t)\cdot\psi_B({\bf r}_{B},\,
t)\approx\Psi({\bf r}_{A},\, {\bf r}_{B},\, t)$. The separability of the full system into its components $A$
and
$B$ is thus ensured, in good approximation, during the scattering process.
\subsection{Mutual scattering of two well localized wave packets (the classical limit-interacting material
points)} Another interesting limiting case is the situation during which we can neglect the quantum extension of
the interacting particles. This will occur when the interaction potential $V_{AB}$ is smooth enough and that the
particles $A$ and $B$ are described by wave packets the extension of which is small in comparison to the
typical lenght of variation of the potential. It is well known that in this regime, when the de Broglie wave
lenghts of the wave packets are large enough, it is consistent to approximate quantum wave mechanics by
its geometrical limit, which is classical mechanics. For instance the quantum differential cross sections
converge in the limit of short wave-lenghts to the corresponding classical cross section. Ehrenfest's theorem
also predicts that when we can neglect the quantum fluctuations, which is the case here, the average motions are
nearly classical and provide a good approximation to the behaviour of the full wave-packet in so far we
consider it as a material point. In this regime, we can in good approximation replace the interaction potential
by the first order term of its Taylor development around the centers of the wave-packets associated to the
particles $A$ and $B$: \begin{displaymath}V_{AB} ({\bf r}_{A}-{\bf r}_{B})\approx V_{AB} (<{\bf r}_{A}>_t-<{\bf
r}_{B}>_t)+{\bf
\nabla}_A V_{AB}(<{\bf r}_{A}>_t-<{\bf r}_{B}>_t)\cdot({\bf r}_{A}-<{\bf
r}_{A}>_t)\end{displaymath}\begin{displaymath}+{\bf \nabla}_B V_{AB}(<{\bf r}_{A}>-<{\bf r}_{B}>_t)\cdot({\bf
r}_{B}-<{\bf r}_{B}>_t).\end{displaymath} Then the evolution equation is in good approximation separable into
the coordinates
${\bf r}_{A},{\bf r}_{B}$ and we have that, when 
$\Psi({\bf r}_{A},\, {\bf r}_{B},\, t=0) = \psi_A({\bf r}_{A},\, t=0)\cdot\psi_B({\bf r}_{B},\, t=0)$, then, at
time
$t$,
$ \Psi({\bf r}_{A},\, {\bf r}_{B},\, t)\approx\psi_A({\bf r}_{A},\, t)\cdot\psi_B({\bf r}_{B},\, t)$ where 
\begin{displaymath}i \hbar\, \partial_t\, \psi_A({\bf r}_{A},\, t) \approx
-{\hbar^2 \over 2m_A} \Delta_{A}\psi_A({\bf r}_{A},\, t)\end{displaymath}\begin{equation} +\,( V_{AB} (<{\bf
r}_{A}>_t>-<{\bf r}_{B}>_t)+{\bf
\nabla}_A V_{AB}(<{\bf r}_{A}>_t-<{\bf r}_{B}>_t)\cdot({\bf r}_{A}-<{\bf r}_{A}>_t))\psi_A({\bf
r}_{A},\, t) \end{equation}
\begin{displaymath}i \hbar\, \partial_t\, \psi_B({\bf r}_{B},\, t) \approx
-{\hbar^2 \over 2m_B} \Delta_{B}\psi_B({\bf r}_{B},\, t)\end{displaymath}\begin{equation} +\,( {\bf
\nabla}_B V_{AB}(<{\bf r}_{A}>_t-<{\bf r}_{B}>_t)\cdot({\bf r}_{B}-<{\bf r}_{B}>_t))\psi_B({\bf
r}_{B},\, t) \end{equation} Note that the Bohmian velocities associated to the particles $A$ and $B$ are
factorisable only when the full state is factorisable. Otherwise, the velocity of a particle depends non-locally
on the location of both particles.

\subsection{Bound states: the Hartree approximation}
When the energy of the full system is negative, we expect that it will remain in a well localised bound state.
When one particle is quite more massive than the rest of the system as is the case with the sun in the solar
system or with the nucleus inside the atom, it is a very good approximation to neglect its recoil and its
extension for all times. Indeed, if we think to the nucleus for instance, its recoil is zero in average over an
orbit, and its Compton wave lenght is very small. Therefore it is consistent in a first approach to reduce the
study of the energy levels of atoms to the study of the energy levels of the electrons that are assumed to
undergo an external central Coulombian potential due to the presence of the nucleus and to factorize the full
wave function into a product of an electronic wave function and of a nuclear one. For sure this approximation is
valid to the extent that we can neglect other degrees of freedom as the nuclear spin and so on which is not
always the case. If moreover we assume that the electronic wave function is itself factorisable, which is
certainly a crude approximation because of the presence of exchange terms due the undistinguishability of the
electrons and because the Coulombian interaction between the electrons is likely to generate entanglement, we are
performing the so called Hartree\footnote{Note that when the Hartree approximation is valid,
particles behave as if they were discernable, and constituted of a dilute, continuous medium
ditributed in space according to the quantum distribution in $\psi_{A(B)}^2(r_{A(B)},t)$, which is
close to the interpretation of the wave-function originally adopted by Schroedinger.} approximation
[6]. Let us consider the Helium atom for instance, and let us neglect the fermionic exchange
contributions, the spins of the electrons and of the nucleus and so on. The time independent
(electronic) Schroedinger equation is then the following:
\begin{equation}E_{AB}\cdot \Psi({\bf r}_{A},\, {\bf r}_{B}) =
(-{\hbar^2 \over 2m_A} \Delta_{A}\, +\,V_A\,-\,  {\hbar^2 \over 2m_B}\Delta_{B}\,+\,V_B)\Psi({\bf r}_{A},\, {\bf
r}_{B})\, +\, V_{AB} ({\bf r}_{A}-{\bf r}_{B})\Psi({\bf r}_{A},\, {\bf
r}_{B}) \end{equation} 
where $V_A$ and $V_B$ represent the external fields (for instance the Coulombian nuclear field), while $V_{AB}$
represents the Coulombian repulsion between the electrons $A$ and $B$. Let us assume that this equation admits a
factorisable solution $\Psi({\bf r}_{A},\, {\bf
r}_{B})$ = $\psi_A({\bf r}_{A})\cdot\psi_B({\bf r}_{B})$; then:

\begin{displaymath}E_{AB}\cdot \psi_A({\bf r}_{A})\cdot\psi_B({\bf r}_{B}) \end{displaymath}\begin{displaymath}=
((-{\hbar^2 \over 2m_A} \Delta_{A}\, +\,V_A)\psi_A({\bf r}_{A}))\cdot\psi_B({\bf
r}_{B})\,+\,
\psi_A({\bf r}_{A})\cdot (-{\hbar^2 \over 2m_B}\Delta_{B}\,+\,V_B)\psi({\bf
r}_{B})\end{displaymath}\begin{equation}\label{def} +\, V_{AB} ({\bf r}_{A}-{\bf r}_{B})\psi_A({\bf
r}_{A})\cdot\psi_B({\bf r}_{B})\end{equation}

Let us now take the in-product of this equation with $\psi_A({\bf r}_{A})$ and multiply the
resulting equation by $\psi_A({\bf r}_{A})$ respectively. We obtain:

\begin{displaymath}E_{AB}\cdot \psi_A({\bf r}_{A})\cdot \psi_B({\bf r}_{B})
\end{displaymath}\begin{displaymath}= \psi_A({\bf r}_{A})\cdot<(-{\hbar^2 \over 2m_A} \Delta_{A}\,
+\,V_A)>_A\cdot\psi_B({\bf r}_{B})\,+\,
\psi_A({\bf r}_{A})\cdot(-{\hbar^2 \over 2m_B}\Delta_{B}\,+\,V_B)\psi({\bf
r}_{B})\end{displaymath}\begin{equation} +\, <V_{AB} ({\bf r}_{A}-{\bf r}_{B})>_A\cdot\psi_A({\bf
r}_{A})\cdot\psi_B({\bf r}_{B})\end{equation}
Similarly, we get that: 
\begin{displaymath}E_{AB}\cdot \psi_A({\bf
r}_{A})\cdot \psi_B({\bf r}_{B})
\end{displaymath}\begin{displaymath}= <(-{\hbar^2 \over 2m_A} \Delta_{B}\, +\,V_B)>_B\cdot\psi_A({\bf
r}_{A})\cdot \psi_B({\bf r}_{B})\,+\,
((-{\hbar^2 \over 2m_B}\Delta_{A}\,+\,V_A)\psi({\bf
r}_{A}))\cdot \psi_B({\bf r}_{B})\end{displaymath}\begin{equation} +\, <V_{AB} ({\bf r}_{A}-{\bf
r}_{B})>_B\cdot\psi_A({\bf r}_{A})\cdot \psi_B({\bf r}_{B})\end{equation}

Let us now sum the two previous equations and substract the eqn.\ref{def}.
We obtain the following consistency condition:
\begin{displaymath}(E_{AB}-<(-{\hbar^2 \over 2m_A} \Delta_{A}\,
+\,V_A)>_A- <(-{\hbar^2 \over 2m_A} \Delta_{B}\, +\,V_B)>_B  )\cdot \psi_A({\bf r}_{A})\cdot\psi_B({\bf r}_{B})
\end{displaymath}\begin{equation} =( <V_{AB} ({\bf r}_{A}-{\bf r}_{B})>_A + <V_{AB} ({\bf r}_{A}-{\bf
r}_{B})>_B-  V_{AB} ({\bf r}_{A}-{\bf r}_{B})  )\cdot\psi_A({\bf r}_{A})\cdot \psi_B({\bf r}_{B})\end{equation}
Equivalently, when the wave-function does not vanish, the following condition must be satisfied:
\begin{equation} V_{AB} ({\bf r}_{A}-{\bf r}_{B})= <V_{AB} ({\bf r}_{A}-{\bf r}_{B})>_A + <V_{AB} ({\bf
r}_{A}-{\bf r}_{B})>_B-  <V_{AB} ({\bf r}_{A}-{\bf r}_{B})>_{AB}  \end{equation}
This is nothing else than the condition \ref{fichtre} in a static form. Note that here it appears to be a necessary condition
, which does not infirm our conjecture that the necessary condition of the lemma is true in the
infinite dimensional case. Reciprocally, it is easy to check that if the eqn.\ref{fichtre} is
satisfied, that the full state is factorisable and that the reduced states of the particles
$A$ and
$B$ are eigenstates of their respective effective Hamiltonians, the full state is eigenstate of the full
Hamiltonian, in agreement with the sufficient condition of the lemma (which is a special case of the
theorem 2). We see thus that the Hartree approximation is valid when the interaction factorises into
the sum of two effective potentials that act separately on both particles, and express the average
influence due to the presence of the other particle (which is not true in general and certainly not
inside the atom). Similarly, in the test-particle limit, the effective potential undergone by the
massive particle is close to zero, and when the heavy particle is well localised, its average,
effective, potential is close to the real potential undergone by the light ``test-particle''. In the
classical limit (material points), the quantum internal structure of the interacting particles can be
neglected and the potential is equivalent to the sum of the effective potentials.

In the three cases, the necessary condition of the lemma is not infirmed, and
its sufficient condition is confirmed. In all the cases, the systems are separated
only in first approximation.

\section{Conclusions and comments}

A conclusion of this work could be: in quantum mechanics to interact means nearly always to
entangle. We showed that real interactions do necessarily generate entanglement (the inverse result,
that it is impossible to generate entanglement without turning on an interaction, is rather trivial).

Considered
so, the degree of entanglement of the universe ought to increase with time, which would indicate some analogy
between entanglement and entropy. Note however that the temporal reversibility of the Schroedinger equation
implies that the degree of entanglement could also decrease in time so that we face a paradox analog to the
famous Loschmidt paradox which emphasises the apparent contradiction between the temporal asymmetry of the second
principle of thermodynamics and the temporal symmetry of fundamental interactions. Obviously, such considerations
are out of the scope of this paper and we invite the interested reader to consult the reference
[7] and references therein.

Beside, it would be worth investigating the generalisation of our results to infinite
dimensional systems. We conjecture that the necessary conditions of our lemma and
of our main theorem are still true when we deal with infinite dimensional systems, as is the case
for the corresponding sufficient conditions and for the theorem 2. 

Let us briefly reconsider the three
situations during which the interaction between two mutually interacting particles (in the three
dimensional, physical space) is entanglement-free, at least in first approximation (see section 3).
These are the test-particle limit (no feedback), the geometrical limit of quantum wave mechanics
(narrow wave-packets) and the Hartree approximation (particles seen as a dilute gas). Each of these
situations has a counterpart in ``classical'' physics: idealised test-particles play an important
role in classical mechanics and in general relativity, the geometrical limit of quantum wave
mechanics is Hamiltonian mechanics, while the image according to which charged particles are
characterised by a spatial distribution (spherical or other) motivated important works in classical
electro-magnetism at the beginning of our century. Entanglement really marks a departure from such
lines of thought, which confirms the deep intuition of Schroedinger, already mentioned in the
introduction, who described entanglement as the characteristic trait of quantum mechanics, ``the one
that enforces its entire departure from classical lines of thought'' [1].

The present work was motivated by the results presented in the references
[8,9]. In these papers it is argued and shown that retrievable, usable quantum information can
be transferred in a scheme which, in striking contrast to the quantum teleportation schemes, requires no external
channel and does not involve the transfer of  a quantum state from one subsystem to the other. Entanglement-free
interaction between two mutually scattering particles (in the three dimensional, physical space) plays a crucial
role in this scheme. The previous remarks suggest that localisation of at least one of the particles
is a necessary ingredient of such protocols for quantum information transfer. For instance, in the
test-particle limit the massive particle is localised while in the classical limit, both particles
are localised. It is easy to show that if at least one of the two interacting particles is not well
localised (bilocated for instance), and that the particles interact through a position-dependent
potential (action at a distance), they are highly likely to end up in an entangled state.

\section*{Acknowledgements}
Sincere thanks to John Corbett (Macquarie's University, Sydney) for his fruitful discussions and
comments. This work originated during my visit at Macquarie's university in March and April 2001.
Support from the Fund for Scientific Research, Flanders, is acknowledged.

\section*{References}

\noindent [1] E. Schroedinger, {\it Discussion of probability relations between separated systems}, Proc.
Cambridge Philos. Soc. {\bf 31}, 555 (1935).

\noindent [2] J. S. Bell, {\it On the EPR paradox}, Physics, {\bf 1}, 165 (1964).

\noindent [3] N. Gisin, {\it Bell's inequality holds for all non-product states}, Phys. Lett. A {\bf 154}, n$^o$
5,6, 201 (1991).

\noindent [4] D. Home and F. Selleri, {\it Bell's
theorem and the EPR paradox}, La Rivista del Nuovo Cimento della Societa Italiana di fisica, {\bf 14}, n$^o$ 9
(1991) p 24. 

\noindent [5] A. Peres, {\it Quantum Theory: Concepts and Methods}, Kluwer Dordrecht (1993) p123.

\noindent [6] L. Landau and E. M. Lifshitz, {\it Non-Relativistic Quantum Mechanics}, Pergamon Press
Oxford (1962) p234.

\noindent [7] J. Gemmer, A. Otte and G. Mahler, {\it Quantum approach to a derivation of the second
law of thermodynamics}, Phys. Rev. Lett. {\bf 86}, 1927 (2001).

\noindent [8] J. Corbett and D. Home, {\it Quantum effects involving interplay between unitary
dynamics and kinematic entanglement}, Phys. Rev. A, {\bf 62}, 062103 (2000).

\noindent [9] J. Corbett and D. Home, {\it Ipso-Information-transfer}, quant-ph/0103146.

\end{document}